\documentclass[12pt]{article}
\usepackage{amsmath}
\usepackage{epsfig}

\begin{document}
\begin{titlepage}
\begin{flushright}
CERN-TH/2003-007
\end{flushright}

\vspace{1.3cm}

\begin{center}
{\large \bf An alternative model of jet suppression at RHIC energies}
\end{center}
\vspace{1cm}
\begin{center}
{Roman Lietava${}^{a}$, J\'an Pi\v{s}\'ut${}^{b}$,
Neva Pi\v{s}\'utov\'a${}^{c}$, and Boris Tom\'a\v{s}ik${}^{b}$}
\end{center}

\vspace{0.3cm}

\begin{center}
{${}^{a}$ {\it School of Physics and Astronomy, 
University of Birmingham, Birmingham B15 2TT, United Kingdom}}
\end{center}

\begin{center}
{${}^{b}$ {\it CERN, Theory Division,  CH-1211 Geneva 23, Switzerland}}
\end{center}

\begin{center}
{${}^{c}${\it Department of Physics, Comenius University,
SK-84248 Bratislava, Slovakia}}
\end{center}

\vspace{1cm}
\begin{center}
February 10, 2003
\end{center}

\vspace{0.7cm}

\abstract%
{We propose a simple Glauber-type mechanism for suppression of 
jet production up to transverse momenta of about 10~GeV/$c$ at RHIC. 
For processes in this kinematic region, the formation time is
smaller than the interval between two successive hard partonic 
collisions and the subsequent collision  influences the 
jet production. Number of jets then roughly scales with the
number of participants. Proportionality to the number of binary 
collisions is recovered for very high transverse momenta. The model 
predicts  suppression of jet production in d+Au collisions at RHIC.}
\end{titlepage}

%%%%%%%%%%%%%%%%%%%%%%%%%%%%%%%%%%%%%%%%%%%%%%%%%%%%%%%%%%%%%%%%%%%%%

It was expected that the yields of high-$p_T$ hadrons from 
nuclear collisions at RHIC would scale with the number of 
nucleon-nucleon interactions at the given value of the impact 
parameter. The observed
spectra are suppressed above $p_T \approx 2\,\mbox{GeV}/c$
with respect to this expectation \cite{PHE1,STAR1,PHE2,PHE3,STAR2}.
This suppression 
is most frequently interpreted \cite{VGP02,WGP95,Wie00,GLV00,HN02,BM02} 
as due to  loss of energy of high-$p_T$ partons in quark-gluon plasma (QGP).
Such a mechanism affects the fragmentation of  jet into 
final-state hadrons but does not influence jet production. A slight 
attenuation of jet production may be due to shadowing  \cite{shad}
but even both these effects together are unable to reproduce the strong 
observed suppression.

In this paper we propose a mechanism which leads to attenuated production 
of jets with $2 \le p_T \le 10\, \mbox{GeV}/c$. We will label this kinematic
region as ``medium'' $p_T$ in order to distinguish it from really hard 
jets at higher energies. Owing to the reduced
production of jets, hadronic yields will be suppressed, as well.

We start with the estimate of the formation time of medium-$p_T$ 
jets at RHIC energies, following \cite{PP97}. Our estimate 
is based on a simple Glauber approach which we take as justified 
for hard constituent partons. The mean
free path of  nucleon in a nucleus at rest is 
$\lambda \approx  2.5\,\mbox{fm}$. In the centre of mass frame of a
nuclear collision at RHIC, the mean free path of an incident 
participating nucleon is Lorentz contracted to about 0.025 fm,
whereas in the SPS energy region it is 0.25 fm. At 
RHIC, the time interval between two successive nucleon-nucleon 
collisions is thus $\Delta t\approx 0.025\,\mbox{fm}/c$. From 
uncertainty relation, a process with  
longitudinal momentum transfer $\Delta p_L$ and energy transfer
$\Delta E$ is materialized within the space interval $\Delta z$ and
time interval $\Delta t$, provided that
\begin{equation}
\Delta p_L > \frac {\hbar}{\Delta z}\, , \qquad
\Delta E > \frac {\hbar}{\Delta t}\, .
\label{eq1}
\end{equation}
According to our argument, processes which do not satisfy this condition 
suffer from interference due to subsequent interactions with following 
incident nucleons or hard partons. The limiting values for RHIC and SPS are 
$\Delta p_L \approx  8\, \mbox{GeV}/c$ 
and $\Delta p_L \approx  0.8 \,\mbox{GeV}/c$, respectively.
Since the $p_T$ of jets comes mainly
from the transfer of longitudinal momentum of partons to the
transverse one, the condition
\begin{equation}
\Delta p_L \approx p_T \ge 8\,  \mbox{GeV}/c\, ,
\label{eq2}
\end{equation}
has to be satisfied if the process is to be finished before the next
nucleon comes to the space-time region where the process develops.
Equations~\eqref{eq1} and \eqref{eq2} are approximations which 
fully exploit the uncertainty relation, so we expect that at RHIC
truly hard processes are those with $\Delta p_L$ larger than 12 or 
perhaps even 15 GeV/$c$. This lead us to label those processes 
with $p_T$ in the region 2--8~GeV/$c$ as ``medium-$p_T$'' ones.

When two jets are produced in a collision
of two nucleons and a third nucleon arrives at the production place
such that Eqs.~\eqref{eq1} and \eqref{eq2} 
are not satisfied, we will assume that the process of jet formation can be 
attenuated. A possible mechanism causing this effect is the screening of 
interaction by colour fields of the third nucleon.

We use the original ``Glauber language'' and formulate our mechanism
in terms of nucleon-nucleon collisions. At RHIC energies one could 
object that softer parton fields of incident nucleons spread longitudinally  
more than the naively calculated Lorentz-contracted nucleon size. They 
overlap and
it is hard to talk about individual nucleons as they are hardly localised.
Hard processes, however, come from interactions of hard partons which are 
well localised within the Lorentz-contracted nucleons. In addition, in
order to make our picture consistent, we will assume that hard processes,
during their formation time, can only be influenced by other {\em hard} partons 
which can be localized within nucleons. In this simple model we ignore
the possible influence of softer partons.

Now, we shall describe our qualitative model 
of medium-$p_T$ jet suppression at RHIC energies. In this paper we
remain at the level of jets and do not calculate
hadronic spectra. In order to calculate them we would have to  use 
fragmentation function which depends on medium in 
which the jets fragment. We defer this to further work.

Technically, our Glauber model is 
constructed in analogy with nuclear absorption of 
$J/\psi$ in heavy-ion collisions \cite{GHRev}. The cross-section
for destruction of jets in the stage of formation by an incident
nucleon is parametrized as
\begin{equation}
\sigma_a =      
\sigma_a(p_T)
=\sigma_0 \left( \frac{1}{1+(p_T/p_{T 0})^2}\right)^2
\label{eq3}
\end{equation}
where $p_{T 0} \approx  8 \, \mbox{GeV}/c$ and $\sigma_0$ is of the order of
a few mb. The parametrization has been chosen in such a way that
the suppression disappears for truly hard jets 
when $p_T$ is  larger than $p_{T 0}$ and for low $p_T$
the absorptive cross-section goes to $\sigma_0$. 
% Otherwise the parametrization is admittedly 
% rather arbitrary.
A motivation for the specific choice of $p_T$-dependence of $\sigma_a$
comes from the modification of  Coulomb scattering in Born 
approximation caused by Debye screening, see also \cite{PP97}.  

%%%%%%%%%%%%%%%%%%%%%%%%%%%%%%%%%%%%%%%%%%%%%%%%%%%%%%%%%%%%%%%%%%%%
\begin{figure}[t]
\begin{center}
   \epsfig{file=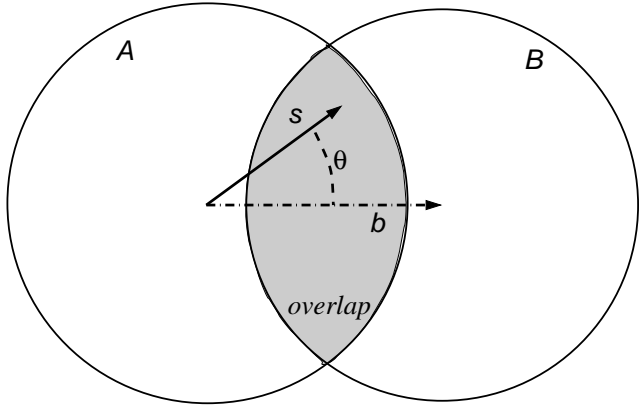,scale=0.38}
\end {center}
\caption{Geometry of non-central collisions.}
\label{fig-geom}
\end{figure}
%%%%%%%%%%%%%%%%%%%%%%%%%%%%%%%%%%%%%%%%%%%%%%%%%%%%%%%%%%%%%%%%%%%%%
In order to simplify further notation we introduce a  shorthand
for the yield of jets with transverse
momenta equal to $p_T$ produced in a collision of nuclei A+B 
at a given value of the impact parameter $b$ 
\begin{equation}
Y_{AB}(p_T,b) = 
\frac{\frac{d\sigma_{AB}}{dp_T^2\, db^2}}{\frac{d\sigma_{AB}}{db^2}}\, .
\label{eq-YAB}
\end{equation}
For normalisation we will use the corresponding yield in proton-proton
collisions
\begin{equation}
\label{eq-Ypp}
Y_{pp}(p_T) = \frac{\frac{d\sigma_{pp}}{dp_T^2}}{\sigma_{pp}}\, .
\end{equation}
We want to determine
%                                                              
%\begin{multline}
\begin{equation}
\frac{Y_{AB}(p_T,b)}{Y_{pp}(p_T)}
=
\int_{\rm overlap} s\, ds \, d\theta
\int_{-L_A}^{L_A} dz_A \, \rho_A\,  
\int_{-L_B}^{L_B} dz_B\, \rho_B\, \sigma_{nn}\,
 F(b,s,\theta,z_A,z_B) \, ,
\label{eq4}
\end{equation}   
where
\begin{equation}
F(b,s,\theta,z_A,z_B)=\exp{[-\sigma_a\rho_A(z_A+L_A)]}
\exp{[-\sigma_a\rho_B(z_B+L_B)]} \, .
\label{eq5}
\end{equation}
For simplicity we work here in  approximation with nuclei as
spheres with constant densities $\rho_A$ and $\rho_B$. The 
first integration in \eqref{eq4} runs over
the overlapping region in non-central collision, see Fig.~\ref{fig-geom}. 
The use of coordinates $s$ and $\theta$ is also explained in that 
Figure. For the non-diffractive nucleon-nucleon cross-section 
we take the value $\sigma_{nn}=$~4~fm$^2$ and $\sigma_a(p_T)$ is given by
eq.~\eqref{eq3}. 
Finally, $2L_A$ and $2L_B$ are lengths of colliding tubes
\begin{equation}
2L_A(s)=2\sqrt{R_A^2-s^2}\, ,\qquad 
2L_B(b,s,\theta)= 2\sqrt{R_B^2-b^2-s^2+2bs\cos\theta}\, ,
\label{eq6}
\end{equation}
in the notation of Fig.~\ref{fig-geom}. 
The coordinates $z_A$ and $z_B$ in eqs.~\eqref{eq4} and \eqref{eq5} specify
positions of those nucleons whose collision led to the production
of two jets. Note that the calculation can
be formulated without accounting for Lorentz contraction of the nuclei 
when determining the values of $L_A$ and $L_B$. In such a case, we have 
to use the standard nuclear density $0.138 \, \mbox{fm}^{-3}$ for
$\rho_A$ and $\rho_B$, in order to proceed in a consistent way.

The number of participating nucleons
$n_{\rm part}$ and the number of binary  collisions
$n_{\rm coll}$ at a given value of $b$ are calculated in the standard 
way
\begin{eqnarray}
\label{eq-part}
n_{\rm part} & = &  \int_{\rm overlap} s\, ds \, d\theta\,
\left ( 2 \rho_A\, L_A(s) + 2\rho_B\, L_B(b,s,\theta)\right )\, ,
\\ 
\label{eq-bin}
n_{\rm coll} & = &  \int_{\rm overlap} s\, ds \, d\theta\,
\sigma_{nn} \, \rho_A \, \rho_B \, L_A(s) \, L_B(b,s,\theta) \, .
\end{eqnarray}

%%%%%%%%%%%%%%%%%%%%%%%%%%%%%%%%%%%%%%%%%%%%%%%%%%%%%%%%%%%%%%%%%%%%
% here we start talking about the results
%%%%%%%%%%%%%%%%%%%%%%%%%%%%%%%%%%%%%%%%%%%%%%%%%%%%%%%%%%%%%%%%%%%%%

Results can be obtained from direct evaluation of 
eqs.~\eqref{eq4}, \eqref{eq-part} and \eqref{eq-bin}, or from 
a simulation of the described model. We chose the latter methode.
%
%%%%%%%%%%%%%%%%%%%%%%%%%%%%%%%%%%%%%%%%%%%%%%%%%%%%%%%%%%%%%%%%%%%%
\begin{figure}[t]
\begin{center}
   \epsfig{file=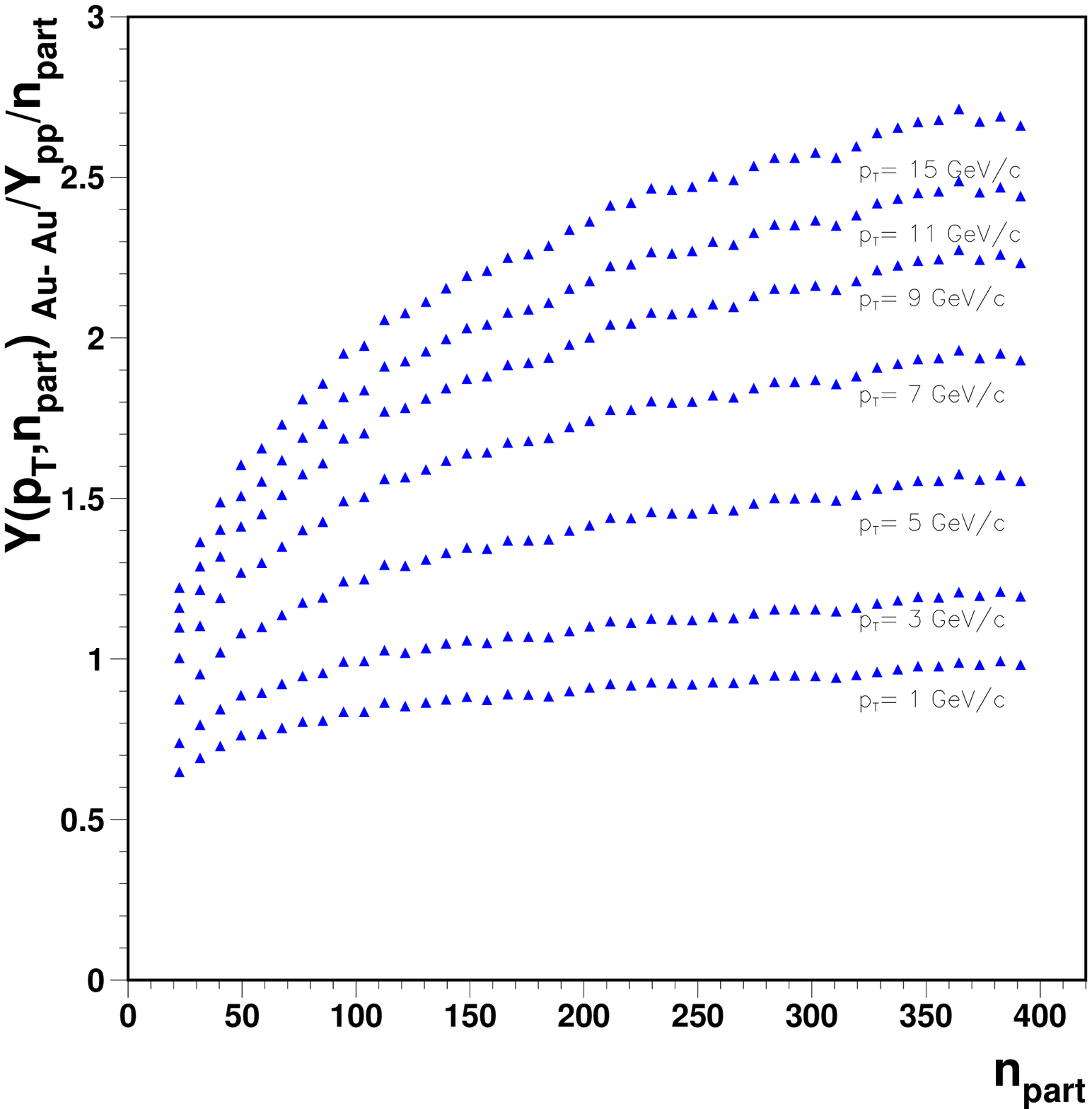,scale=0.52}
\end{center}
\caption{The ratio $Y_{\rm Au+Au}(p_T,b)/[n_{\rm part}(b) \, 
Y_{pp}(p_T)]$ 
plotted as a function of $n_{\rm part}(b)$ for $p_T= 1\,\mbox{GeV}/c$, 
$p_T= 3 \, \mbox{GeV}/c$,
$p_T= 5 \, \mbox{GeV}/c$, $p_T= 7 \, \mbox{GeV}/c$ and  
$p_T=9\, \mbox{GeV}/c$ in Au+Au interactions at
RHIC. Parameters are given in the text, $\sigma_0= 8\,\mbox{mb}$.}
\label{fig1}
\end{figure}
%%%%%%%%%%%%%%%%%%%%%%%%%%%%%%%%%%%%%%%%%%%%%%%%%%%%%%%%%%%%%%%%%%%%%
%
%%%%%%%%%%%%%%%%%%%%%%%%%%%%%%%%%%%%%%%%%%%%%%%%%%%%%%%%%%%%%%%%%%%%
\begin{figure}[t]
\begin{center}
   \epsfig{file=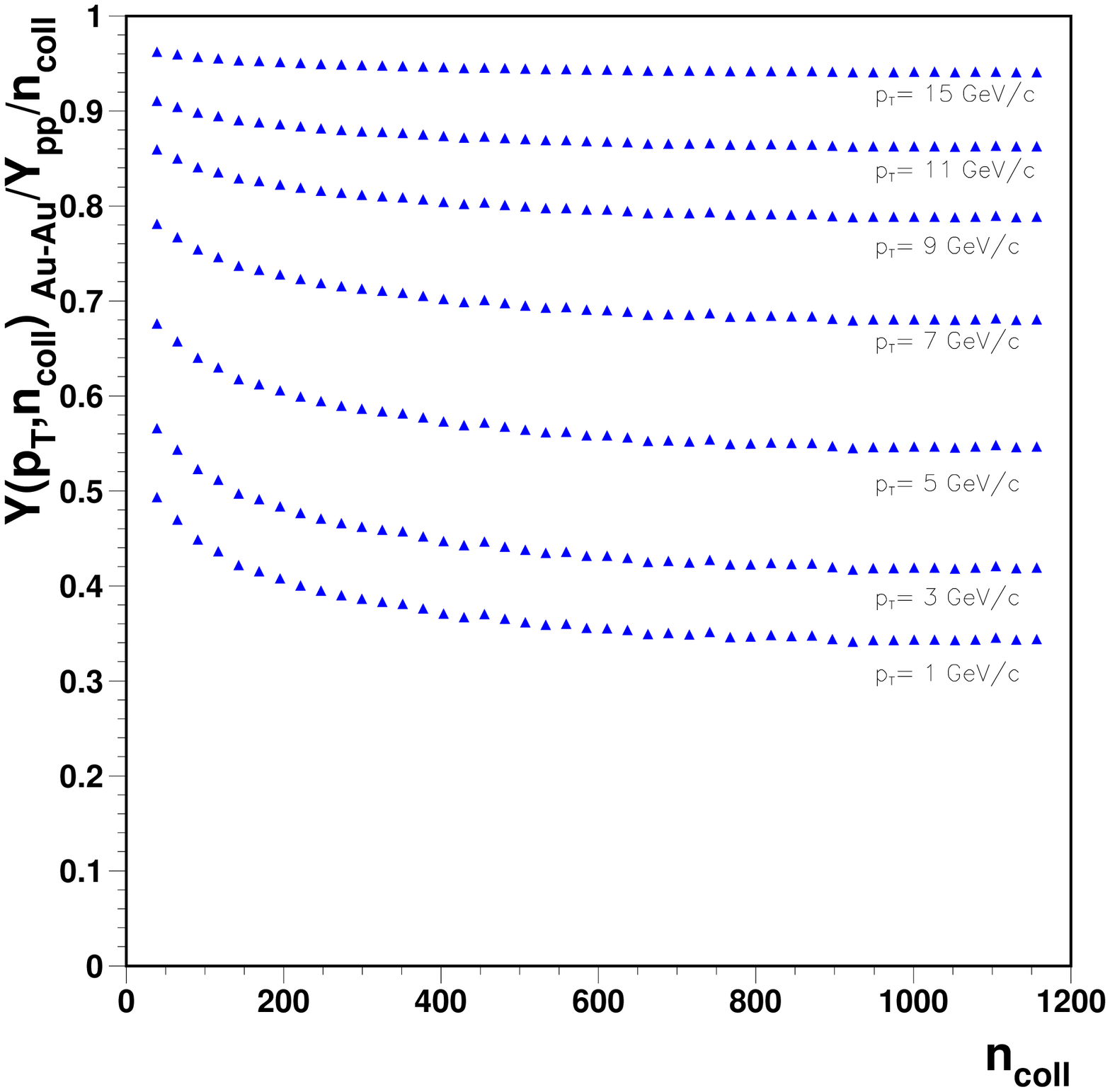,scale=0.52}
\end{center}
\caption{The ratio 
$Y_{\rm Au+Au}(p_T,b)/[n_{\rm coll}(b) \, Y_{pp}(p_T)]$ 
plotted as a function of $n_{\rm coll}(b)$ for $p_T= 1\,\mbox{GeV}/c$, 
$p_T= 3 \, \mbox{GeV}/c$,
$p_T= 5 \, \mbox{GeV}/c$, $p_T= 7 \, \mbox{GeV}/c$ and  
$p_T=9\, \mbox{GeV}/c$ in Au+Au interactions at
RHIC. Parameters are given in the text, $\sigma_0= 8\,\mbox{mb}$.}
\label{fig2}
\end{figure}
%%%%%%%%%%%%%%%%%%%%%%%%%%%%%%%%%%%%%%%%%%%%%%%%%%%%%%%%%%%%%%%%%%%%%
%
%%%%%%%%%%%%%%%%%%%%%%%%%%%%%%%%%%%%%%%%%%%%%%%%%%%%%%%%%%%%%%%%%%%%
\begin{figure}[t]
\begin {center}
   \epsfig{file=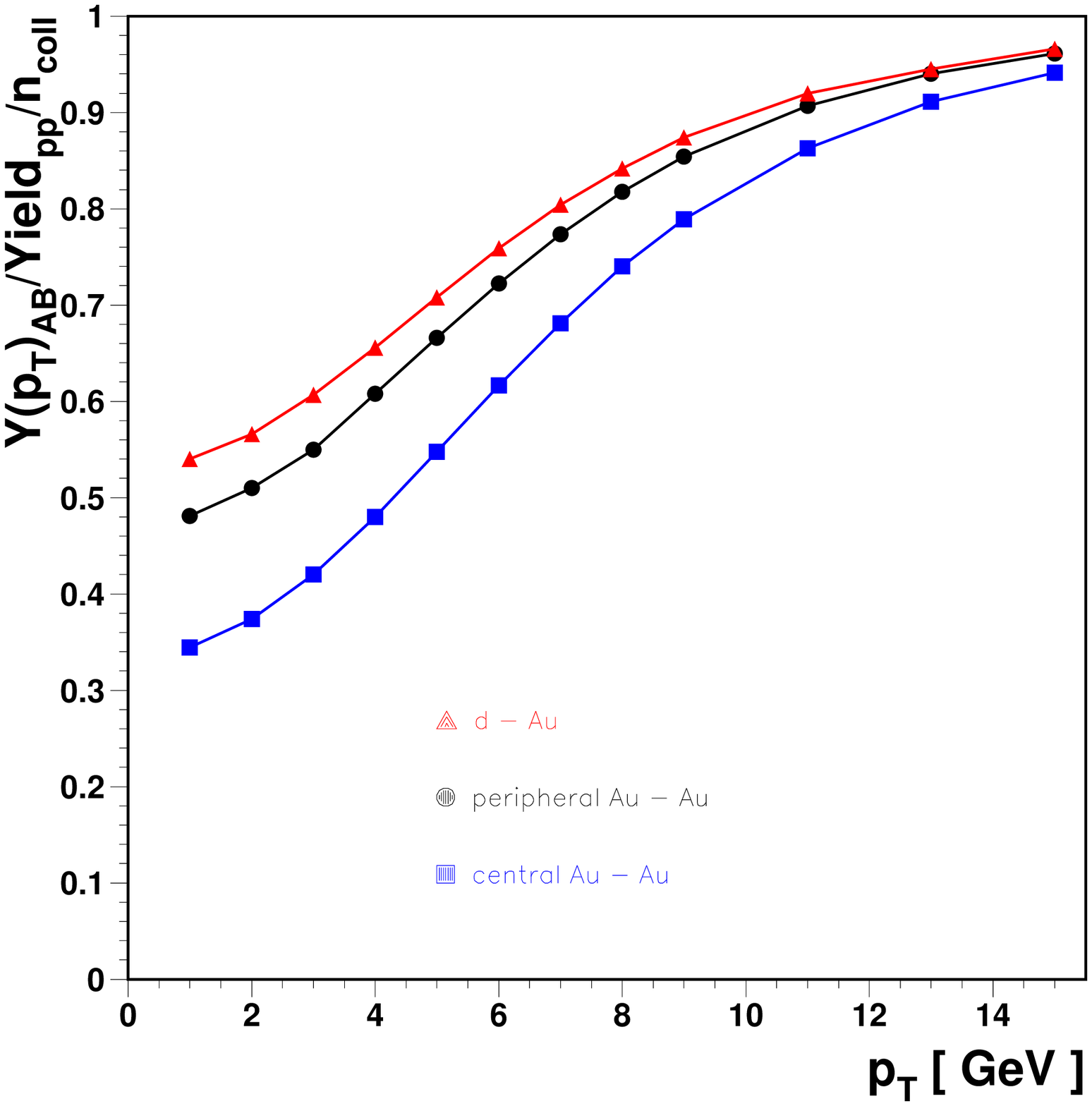,scale=0.52}
\end {center}
\caption{
The ratio $Y_{\rm AB}(p_T)/[n_{\rm coll}\, Y_{pp}(p_T)]$ 
as a function of $p_T$. Results are shown for three different
colliding systems at RHIC: central Au+Au collisions
(0-5\% of the total cross section), peripheral (60-80\%) Au+Au collisions,
and d+Au collisions. We set $\sigma_0= 8\,\mbox{mb}$.}
\label{fig3}
\end{figure}
%%%%%%%%%%%%%%%%%%%%%%%%%%%%%%%%%%%%%%%%%%%%%%%%%%%%%%%%%%%%%%%%%%%%%
%
In Fig.~\ref{fig1} we test the scaling of jet production with the 
number of participants. This regime corresponds to a constant curve
in Fig.~\ref{fig1} and is realized as $p_T \to 0$. For 
increasing $p_T$ the scaling with number of binary collisions
is recovered. This is seen in Fig.~\ref{fig2}: for $p_T > p_{T 0}$
the curves are close to the asymptotic value of 1.

Scaling with the number of participants at low $p_T$ comes from
the fact that most of the jets produced in the volume are suppressed
by nuclear absorption and only those originating from the rear parts
of the colliding nuclei survive. In this way, the jet production is
close to the surface effect.
With growing impact parameter the volume decreases faster than 
the surface, and so production from the surface makes up a bigger
part of the total yield. Thus peripheral collisions
should come to the regime of scaling with number of binary collisions
faster. This is indeed seen in Fig.~\ref{fig3}.

The described mechanism also leads to suppression of {\em jet} 
production in proton-nucleus or deuteron-nucleus collisions.
Our prediction for d+Au is shown in Fig.~\ref{fig3}. 
Data from this collision system should be available soon.

We should stress again, that in this paper we only calculate how the jet 
production is suppressed and do not include fragmentation
into hadrons. Our results thus show qualitative features
which will be seen in the hard hadronic spectra, but cannot 
be directly compared with measured spectra of high-$p_T$ hadrons.

%%%%%%%%%%%%%%%%%%%%%%%%%%%%%%%%%%%%%%%%%%%%%%%%%%%%%%%%%%%%%%%%%%

Before concluding let us add a few comments:
\begin{enumerate}
\item 
Suppression of hard processes in p+A  interactions based on 
shadowing and incident 
parton energy loss in nuclear matter has been discussed
in \cite{Rau,Kop1,Kop2}. In this approach the shadowing is 
understood as destructive interference derived from the 
space-time properties of nuclear interaction which is of a similar 
origin as the suppression considered in our model. The effect 
of incident parton energy loss while traversing nuclear matter before 
the hard collision which leads to jet production is not included
in the present version of the model, but can be added in the 
future.
\item %
We want to stress that the model described above cannot 
explain  the disappearance of opposite-side jet.  
If fully confirmed, this effect will give evidence in favour
of jet suppression by hard parton energy loss in quark-gluon plasma. 
Jet suppression by
QGP is not included in our model, but it can be added later.
\item 
As shown in \cite{Hwa}, nuclear absorption in p+A and A+B
interactions leads to very similar results as gluon depletion
due to energy loss of gluons in nuclear matter. Since absorption,
as shown in Figs.~\ref{fig1} and \ref{fig2} can convert the 
``volume'' effect
to the ``surface'' one, it is likely that gluon energy loss
in nuclear matter could lead to similar results.
\item 
For SPS energy region, the value of $p_{T0}$ is about 0.8 GeV/c.
Hence, the production of particles with $p_T\le $ 0.8 GeV/c
is suppressed and scales rather with $n_{\rm part}$ than with 
$n_{\rm coll}$,
in accord with the venerable wounded nucleon model \cite{WNM} and 
with the data \cite{scaling}.
\end{enumerate}

We conclude that nuclear absorption can attenuate production 
of medium-$p_T$ hadrons. Their yield then scales roughly with 
the number of participants. 

\paragraph{Note added.}
After having submitted this paper we have learned that jet quenching
in Au-Au data at RHIC has been obtained by Kharzeev, Levin and
McLerran \cite{KLL} by parton saturation \cite{GLR,MQ,LR}
and that they have also predicted jet quenching in d-Au interactions,
see also  \cite{KLN}.

The approach of Refs.\ \cite{KLL,KLN} may seem at the first sight as quite different
from what we have proposed above but in fact the two are rather similar.
In Refs.\ \cite{KLL,KLN} jets are quenched because due to parton saturation
not all gluons in the nucleus can independently scatter off the gluons
in the other nucleus.
In our approach  gluon--gluon scattering at low
$p_T^2\approx Q^2 \leq Q_s^2\approx p_0^2$ is
less efficient due to the space-time limitations imposed upon the interaction.
Perhaps the two approaches are two ways of describing similar effects.

We have also realized that G.Papp {\it et al.} \cite{Papp02} have shown that
the ability of a proton to give rise to multiple hard collisions
in pA interactions even in the CERN SPS energy range decreases
with the number of interactions in a nucleus.

We have also learned about the interesting centrality scaling
\cite{SBKLV,HWA03} in the pion $p_T$-spectra at RHIC,
which is becoming a challenge for model builders.

\paragraph{Acknowledgements.}
One of us (JP) is
indebted to the CERN Theory Division for the hospitality
extended to him. We would like to thank D. Kharzeev, C. Salgado, and
U. Wiedemann for usefull discussions and to P. Jacobs, G. Papp and W. Zajc for
interesting comments.
The work of NP and JP has also been supported by the grant of the Slovak
Ministry of Education No VEGA V2F13.

\end{document}